\documentclass[prd,twocolumn,showpacs,nofootinbib]{revtex4}
\usepackage{graphicx}
\usepackage{amsmath}
\usepackage{bm}
\begin{document}

\newcommand{\Rvir}{$R_{\mathrm{vir}}$}
\newcommand{\Rvire}{R_{\mathrm{vir}}}
\newcommand{\Mvir}{$M_{\mathrm{vir}}$}
\newcommand{\Mvire}{M_{\mathrm{vir}}}
\newcommand{\tdyn}{$t_{\mathrm{dyn}}$}
\newcommand{\tdyne}{t_{\mathrm{dyn}}}
\newcommand{\vk}{$v_{\mathrm{k}}$}
\newcommand{\vke}{v_{\mathrm{k}}}
\newcommand{\vvir}{$v_{\mathrm{vir}}$}
\newcommand{\vvire}{v_{\mathrm{vir}}}
\newcommand{\rhose}{\rho_{\mathrm{s}}}
\newcommand{\rse}{r_{\mathrm{s}}}
\newcommand{\Deltave}{\Delta_{\mathrm{v}}}
\newcommand{\rhoue}{\rho_{\mathrm{u}}}
\newcommand{\rhe}{r_{\mathrm{h}}}

\title{Dark-Matter Decays and Self-Gravitating Halos}
\author{Annika H.\ G.\ Peter}
\email{apeter@astro.caltech.edu}
\affiliation{California Institute of Technology, Mail Code 249-17, 
  Pasadena, California 91125, USA}
\author{Christopher E. Moody}
\affiliation{Department of Physics, University of California, Santa Cruz, California 95064, USA}
\author{Marc Kamionkowski}
\affiliation{California Institute of Technology, Mail Code 350-17, 
  Pasadena, California 91125, USA}

\date{\today}

\begin{abstract}
We consider models in which a dark-matter particle decays to a
slightly less massive daughter particle and a noninteracting
massless particle.  The decay gives the daughter particle a
small velocity kick.  Self-gravitating dark-matter halos that
have a virial velocity smaller than this velocity kick may be
disrupted by these particle decays, while those with larger
virial velocities will be heated.  We use numerical simulations
to follow the detailed evolution of the total mass and density
profile of self-gravitating systems composed of particles that
undergo such velocity kicks as a function of the kick speed
(relative to the virial velocity) and the decay time (relative
to the dynamical time).  We show how these decays will affect
the halo mass-concentration relation and mass function.
Using measurements of the halo mass-concentration
relation and galaxy-cluster mass function to constrain the
lifetime--kick-velocity parameter space for decaying dark
matter, we find roughly that the observations rule out the
combination of kick velocities greater than 100~km~s$^{-1}$ and
decay times less than a few times the age of the Universe.
\end{abstract}

\maketitle

\section{Introduction}\label{sec:intro}

There is a good consensus from many types of observations that
dark matter makes up $\sim 25\%$ of the mass-energy density of
the Universe
\cite{kessler2009,breid2009,vikhlinin2009b,komatsu2010,rozo2010}.
However, the nature of dark matter is unknown.  The most popular
class of candidates is the weakly interacting massive
particle (WIMP), since such particles may be produced
thermally in the early Universe at the right abundance and with
behavior consistent with a large set of cosmological
observations
\cite{jungman1996,appelquist2001,cheng2002,servant2002,hubisz2005}.
The canonical WIMP is electrically neutral, and it is stable.
Once produced in the early Universe, it interacts with itself
and with ordinary matter only gravitationally.  If the WIMP is
the dark matter, then the dark halos of galaxies and galaxy
clusters are described by a collisionless gas of
self-gravitating WIMPs.

However, there is both theoretical and observational room for
other types of dark-matter candidates.  While observations are
\emph{consistent} with WIMPs, the observations do not
\emph{require} the dark matter to be WIMPs.  Moreover, some
observations---e.g., of the inner mass distribution in galaxies
and/or the abundance subhalos in dark-matter halos
\cite{moore1999a}---have inspired theoretical searches for dark-matter candidates with properties beyond
those of the canonical WIMP.  There is a large amount of literature on
dark-matter particles with additional physical properties beyond
those of the canonical stable collisionless WIMP.  Examples
include (but are by no means limited to) particles with small
electric charges \cite{Davidson:2000hf} or dipoles \cite{Sigurdson:2004zp},
self-interacting particles \cite{spergel2000}, or
particles with long-range forces
\cite{Gubser:2004uh,Kesden:2006vz,Kesden:2006zb,ackerman2009,feng2009}
(see Refs.~\cite{D'Amico:2009,feng2010} for recent reviews). 

In the spirit of these lines of investigation, we consider in
this work a neutral dark-matter particle $X$ of mass $M_X$ that
decays with lifetime $\tau$ to a slightly less massive neutral
particle $Y$ of mass $M_Y =M_X(1-\epsilon)$ with $\epsilon \ll1$
and an effectively massless particle $\zeta$ which is itself
assumed to be noninteracting.  We imagine that such a scenario
may arise in some implementations of models of inelastic dark
matter \cite{TuckerSmith:2001hy}.

When the $X$ particle decays, the daughter $Y$ particle receives
a nonrelativistic velocity kick $\vke \simeq \epsilon c$.
Now suppose that these particles make up a self-gravitating halo
of virial velocity $\vvire$.  If $\vke \ll \vvire$, then the
halo may be heated slightly, and its mass distribution thus
rearranged slightly, by the velocity kicks imparted to the $Y$
particles once most of the $X$ particles have decayed.  If, on
the other hand, $\vke \gg \vvire$, then these halos will be
completely disrupted after most of the $X$ particles have
decayed.  There should thus be no halos with $\vvire \ll
\vke$ if these particles decay with lifetimes $\tau$ small
compared with the age $t_\mathrm{H}$ of the Universe, and this has been
postulated as a possible explanation for the low number of dwarf
galaxies relative to that expected from dissipationless
dark-matter simulations \cite{sanchez2003,abdelqader2008}.  Alternatively, if
$\vke \gg \vvire$ but $\tau \gg t_\mathrm{H}$, then only a small fraction
($\sim \tau/t_\mathrm{H}$) of the halo particles will be kicked out of
the halo.  The resulting halo mass and mass distribution may
then be affected slightly without being completely disrupted.
The canonical-WIMP halo is, of course, recovered in the limits
$\tau \to \infty$ and/or $\epsilon \to 0$.

In this paper, we report on simulations of self-gravitating
halos with decaying particles.  While the results of
these decays can be understood in the limits $\tau\gtrsim
t_\mathrm{H}$ and $\vke\gg \vvire$ \cite{peter2010} using the
adiabatic-expansion model \cite{flores1986,cen2001},
detailed evolution of the halo over the full range of the
$\tau$-$\vke$ parameter space requires numerical simulation.
We use the simulations and observations of the galaxy-cluster mass function and the halo
mass-concentration relation to constrain the decay parameter space.  Our central results are presented
in Fig.~\ref{fig:summary}, which shows that (roughly) the
combination of decay times less than a few times the age of the Universe and
kick velocities greater than 100~km~s$^{-1}$ are ruled out.

The outline of the paper is as follows.  In Sec.~\ref{sec:sims},
we describe our simulations.  In Sec.~\ref{sec:results}, we
characterize the evolution of dark-matter halos as a function of
the decay parameters using the simulations, and show how the
simulations, in conjunction with observational and theoretical (in the
context of WIMPs) determinations of the  cluster mass function
and the mass-concentration relation, allow us to constrain the
decay parameter space.  In Sec.~\ref{sec:discussion}, we discuss
various aspects of our findings, including resolution effects
(with an eye towards the requirements for cosmological
simulations) and observational biases.  We summarize our work in
Sec.~\ref{sec:conclusion}.

\section{Simulations}\label{sec:sims}

We simulate isolated equilibrium dark-matter halos using the
parallel $N$-body code GADGET-2 in its Newtonian,
noncosmological mode \cite{springel2001a,springel2005}.  We
have modified GADGET-2 in the following way to handle decays.
The maximum time step $\Delta t_{\mathrm{max}}$ allowed in the
simulations is chosen such that the probability for a decay
within that maximum time step is well-described by
$P_\mathrm{max} = \Delta t_{\mathrm{max}}/\tau \ll 1$.  At each
time step, there is a Monte-Carlo simulation of decays, with the
decay probability being $P = \Delta t / \tau$ for each particle.
If a particle is designated for decay, it receives a kick speed
\vk~ in a random direction, and is flagged to ensure that the
particle cannot decay again.  Because we assume the mass
difference is negligible throughout the simulations ($\epsilon
\lesssim 10^{-3}$), we maintain the mass of the particles; the
change in the kinetic energy of the particle due to the kick is
much greater than the change to the potential energy due to the
decrease in particle mass.  The massless $\zeta$ particle escapes from the halo and is not of interest in this calculation.

Our halos are spherically symmetric, and the density $\rho(r)$
as a function of galactocentric radius $r$ is taken to be the
Navarro-Frenk-White (NFW) form \cite{navarro1996,navarro1997},
\begin{eqnarray}\label{eq:rho}
	\rho(r) = \frac{\rho_\mathrm{s}}{\displaystyle
        \frac{r}{r_\mathrm{s}}\left(1+\frac{r}{r_\mathrm{s}}\right)^{2}},
\end{eqnarray}
where $\rho_\mathrm{s}$ is the scale density, and $r_\mathrm{s}$
is the scale radius, such that the halo concentration is
\begin{eqnarray}\label{eq:c}
	c = \frac{\Rvire}{\rse},
\end{eqnarray}
where \Rvir~is the virial radius, or size, of the halo.  We define virial quantities with respect to the spherical
top-hat overdensity, such that the virial mass
\begin{eqnarray}
     \Mvire = \frac{4\pi}{3}\Delta_\mathrm{v}\rho_\mathrm{m} \Rvire^3,
\end{eqnarray}
where $\rho_\mathrm{m}$ is the mean matter density of the Universe,
and $\Delta_\mathrm{v} = (18\pi^2 - 39x - 82x^2)/\Omega_\mathrm{m}(z)$ in
a $\Lambda$CDM Universe, where $x = \Omega_\mathrm{m}(z) - 1$ and
$\Omega_\mathrm{m}$ is the fraction of the critical density
$\rho_\mathrm{c}$ in the form of matter ($\rho_\mathrm{m} =
\Omega_\mathrm{m} \rho_\mathrm{c}$) \cite{bryan1998}.

The initial velocity distribution must be taken to be consistent
with the mass distribution so that the effects of decay are not
confused with relaxation of an initially nonequilibrium halo to equilibrium.
We take the velocity ellipsoid to be isotropic, a reasonable
first approximation to dark-matter halos \cite{diemand2009}.
This allows us to use Eddington's formula,
\begin{eqnarray}\label{eq:eddington}
	f(\mathcal{E})=\frac{1}{\sqrt{8}\pi^{2}}\left[\int_{0}^{\mathcal{E}}\frac{d^{2}\rho}{d\Psi^{2}}\frac{d\Psi}{\sqrt{\mathcal{E}-\Psi}}+\frac{1}{\sqrt{\mathcal{E}}}\left(\frac{d\rho}{d\Psi}\right)_{\Psi=0}\right] 
\end{eqnarray}
to relate the particle distribution function (DF)
$f(\mathcal{E})$ as a function of the negative energy
$\mathcal{E}$ ($\mathcal{E} > 0$ for all particles bound to the
halo) to known functions, the density and gravitational
potential profile $\Psi$. 

We cannot use Eq. (\ref{eq:rho}) in Eq. (\ref{eq:eddington}) for
arbitrary $r$; the mass of a halo corresponding to a density
profile given by Eq. (\ref{eq:eddington}) is formally infinite.
Thus, we must truncate the mass profile of the halo, but with
care such that the distribution function is everywhere
positive-definite.  We use the truncation scheme advocated by
Kazantzidis et al. (2004) \cite{kazantzidis2004},
\begin{eqnarray}
	\rho(r)=\begin{cases}
	\frac{\displaystyle \rho_\mathrm{s}}{\displaystyle \frac{r}{r_\mathrm{s}}\left(1+\frac{r}{r_\mathrm{s}}\right)^{2}} & r<r_{c}\\
	\frac{\displaystyle \rho_\mathrm{s}}{\displaystyle \frac{r}{r_\mathrm{s}}\left(1+\frac{r}{r_\mathrm{s}}\right)^{2}}\left(\displaystyle \frac{r}{r_{c}}\right)^{\alpha}e^{-(r-r_{c})/r_{d}} & r\geq r_{c}.\end{cases}
\end{eqnarray}
The  density profile has been exponentially truncated at the
cut off radius $r_\mathrm{c}$ and with the parameter
$r_\mathrm{d}$ controlling the sharpness of the transition.
Continuity of the function and its logarithmic slope demand that
\begin{eqnarray}
     \alpha=-\frac{1+3r_\mathrm{c}/r_\mathrm{s}}
     {1+r_\mathrm{c}/r_\mathrm{s}} + \frac{r_\mathrm{c}}{r_\mathrm{d}}.
\end{eqnarray}

To set the initial positions and velocities of the particles, we
sample the DF using an acceptance-rejection method to find the
initial radii and speeds \cite{press1992}.  The radius and
speed are then isotropically mapped to three spatial
coordinates.

\subsection{Parameters}

We begin every simulation with identical initial conditions,
varying only the concentration $c$ of the NFW profile, particle
lifetimes, and kick velocities.  We simulate $\Mvire =
10^{12}M_\odot$ halos only, but we can extrapolate our results
to other halo masses by considering the structural changes to
the halos as a function of the decay parameters with respect to
virial parameters, as discussed in
Sec.~\ref{sec:results:theo}.  To determine the virial spherical
top-hat overdensity $\Delta_\mathrm{v}$, and hence the virial
radius, we use a cosmology with $h = 0.7$, $\Omega_\mathrm{m} =
0.3$, and $\Omega_\Lambda = 0.7$.  We perform a set of 50
simulations, with $\vke = 10,$ 50, 100, 200, and 500 km s$^{-1}$
and $\tau = 0.1,$ 1, 10, 50, and 100 Gyr, and with $c=5$ and $c
= 10$.  Since the virial speed of the halo is $\vvire \equiv
\sqrt{G\Mvire/\Rvire} \approx 130\hbox{ km s}^{-1}$ using a
spherical top-hat overdensity definition of virial parameters,
the kick speeds were chosen to bracket the virial value.  The
typical crossing time of a particle in the halo is $\sim 500$ Myr,
and the virial time $t_\mathrm{vir} = \Rvire / \vvire$ is a few
Gyr, meaning that our choices for $\tau$ bracket the typical
dynamical time scales of particles in the halo.  We choose $c=5$
and $c=10$, since these are typical $\Lambda$CDM concentrations
for galaxies, groups, and clusters.  We use observations of such
systems to set constraints on the decay parameter space in
Sec.~\ref{sec:results:obs}. 

The fiducial simulations used for our analysis in
Sec.~\ref{sec:results} contain $1.5\times 10^6$ particles total,
or $10^6$ particles within the virial radius at the start of the
simulations.  To test the convergence of the simulations, we run
the same simulations over again but with an order of magnitude
fewer particles.  The convergence
tests are further discussed in Sec.~\ref{sec:discussion:res}.

We set GADGET-2 parameters, such as the softening, maximum time
step, and the force accuracy criterion, according to the recommendations
of Power et al. (2003) \cite{power2003}.  For the fiducial
simulations, the softening length is set to 1.3 kpc, and it is set
to 3.3 kpc in the low-resolution simulations.  For most of the
simulations, the maximum allowed time step is set to $\Delta
t_\mathrm{max} = 10$ Myr, in accordance with Power et
al. (2003).  However, for simulations with $\tau = 100$ Myr and
$\tau = 1$ Gyr, $\Delta t_\mathrm{max} = 1$ Myr such that
$P_\mathrm{max} \ll 1$.  We run the simulations for 10 Gyr.

We also performed simulations without decay to check that our
equilibrium initial conditions were indeed in equilibrium, and
to determine the extent of numerical relaxation at the halo
centers.  We found that the halos were stable over the 10 Gyr duration of
the simulations, and that numerical relaxation ceased to be
important for $r > 4$ kpc for the fiducial simulations and $r >
10$ kpc for the low-resolution simulations.

\section{Results}\label{sec:results}

In Sec.~\ref{sec:results:theo}, we present and summarize the
results of our simulations.  We show how the halo mass, NFW
concentration, and NFW goodness of fit depend on the decay
parameters $\vke$ and $\tau$, initial concentration, and time.
In Sec.~\ref{sec:results:obs}, we show how the results of the
simulation can be used in combination with the observed galaxy-cluster mass function and the halo mass-concentration relation
to constrain the decay parameter space.

\subsection{Halo profiles as a function of $v_k$ and
$\tau$}\label{sec:results:theo}

\begin{figure}
\begin{centering}
	\includegraphics[width=0.45\textwidth]{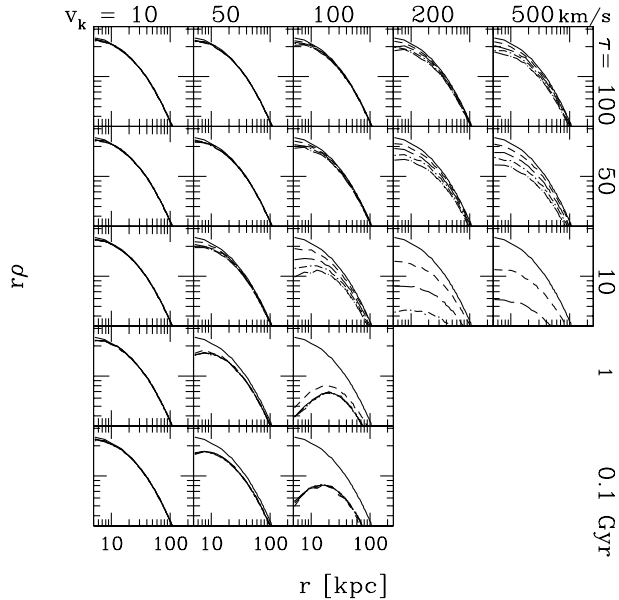}
\end{centering}
\caption{\label{fig:rrho}The halo dark-matter density profile
times the radius, as a function of radius and time.  The lines
correspond to different time slices in the simulation.  The
solid line represents the density at the beginning of the
simulations; the dashed line is the density 2.5 Gyr after the
simulation start; the long-dashed line is that after 5 Gyr; the
dot-dashed line is the density profile after 7.5 Gyr; and the
long-dash-dotted line represents the density profile after 10
Gyr.  Density profiles for $\vke \ge 200\hbox{ km s}^{-1}$ and
$\tau \le 1$ Gyr are not shown because the halos are destroyed
within a few Gyr.}
\end{figure}

We present the halo density profiles, multiplied by the radius
to highlight differences, as a function of \vk, $\tau$, and time
in Fig.~\ref{fig:rrho}.  We show the density profiles for halos
with initial $c=5$ since the qualitative features of the $c=5$
and $c=10$ halos are similar, and omit figures for $\tau = 0.1$
and 1 Gyr with $\vke = 200\hbox{ and } 500\hbox{ km s}^{-1}$
since the halos are effectively destroyed within a few Gyr.  The
higher lines in each subplot correspond to earlier times, and
the lines go (top to bottom) $t = 0$, 2.5, 5, 7.5, and 10 Gyr
after the start of the simulation.

There are other interesting features of note in Fig.~\ref{fig:rrho}.
First, we see that for small \vk~relative to the virial speed,
there is little change to the dark-matter profile.  This is
expected, as such small \vk~leads to neither significant mass
loss nor kinetic energy injection in the halo.  However, for
$\vke = 50\hbox{ km s}^{-1}$ ($\vke / \vvire \sim 0.4$), changes
to halo structure begin to become significant for $\tau \lesssim
10$ Gyr.  In general, changes to halo mass and halo structure
are most extreme for large \vk~and small $\tau$.  Changes are
abrupt when $\tau$ is small, although it takes $\sim 5$ Gyr for
the halos to reach equilibrium if $\tau \le 1$ Gyr.

\begin{figure}
\begin{centering}
	\includegraphics[width=0.45\textwidth]{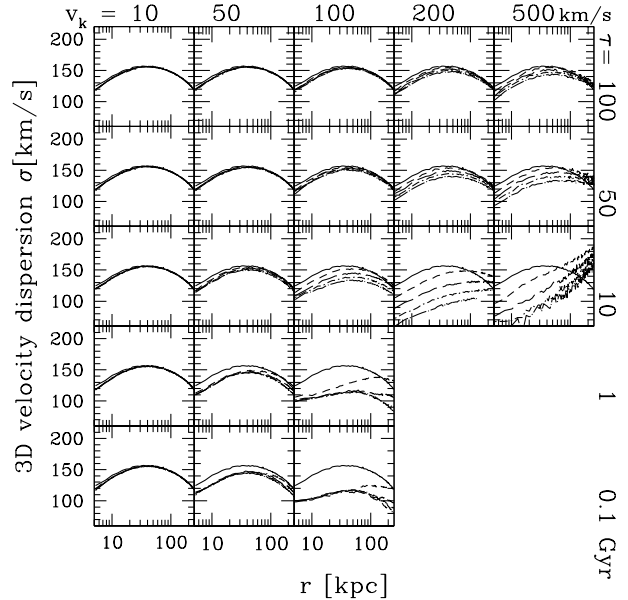}
\end{centering}
\caption{\label{fig:sigma}The three-dimensional velocity dispersion as a function of radius and time, tiled according to \vk~and $\tau$.  The lines
correspond to the same time slices as in Fig. \ref{fig:rrho}.}
\end{figure}

\begin{figure}
\begin{centering}
	\includegraphics[width=0.45\textwidth]{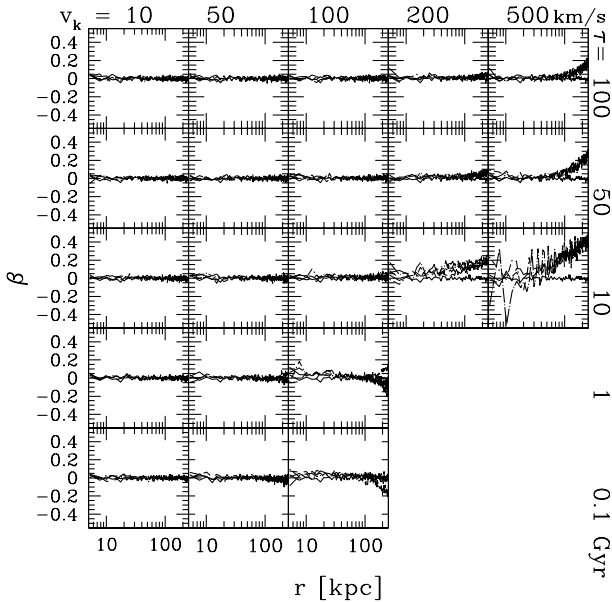}
\end{centering}
\caption{\label{fig:beta}The velocity anisotropy $\beta = 1 - (\sigma_\theta^2 + \sigma_\phi^2)/2\sigma_r^2$ as a function of radius and time, tiled according to \vk~and $\tau$. The lines
correspond to the same time slices as in Fig. \ref{fig:rrho}.}
\end{figure}

We see similar trends in the velocity data, which we present in Figs. \ref{fig:sigma} and \ref{fig:beta}.  In Fig. \ref{fig:sigma}, we show the velocity dispersion profiles 
\begin{eqnarray}
	\sigma \equiv \left[ \sigma_r^2 + \sigma_\theta^2 + \sigma_\phi^2 \right]^{1/2}
\end{eqnarray} 
as a function of decay parameters and time, and in Fig. \ref{fig:beta}, we show the velocity anisotropy
\begin{eqnarray}
	\beta \equiv 1 - \frac{\sigma_\theta^2 + \sigma_\phi^2}{2\sigma_r^2},
\end{eqnarray}
where $\sigma_r$, $\sigma_\theta$, and $\sigma_\phi$ are the velocity dispersions in spherical coordinates.  We set up the initial conditions such that $\sigma_r = \sigma_\theta = \sigma_\phi$ (such that $\beta = 0$) and given the equilibrium initial conditions, the velocity dispersion is initially equivalent to the rms velocities.  Even at late times in the cases of small $\tau$ or large \vk, the rms velocities are only slightly different from the velocity dispersions.

As expected, the velocity dispersion profiles $\sigma(r)$ decrease in normalization as more halo particles decay, since the negative heat capacity of a self-gravitating system ensures that an injection of kinetic energy due to decays results in an overall decrease in the equilibrium kinetic energy of the halo (cf. \cite{peter2010}).  The most extreme drops in the velocity dispersion occur for small $\tau$ or large \vk~for which the kinetic energy injected into the halos due to decays is large.  The velocity dispersion profile quickly settles to a new equilibrium for $\tau \lesssim 1$ Gyr.  When both \vk~is large and $\tau$ is relatively small, the velocity dispersion increases as a function of time at large $r$.  These regions are no longer within the virial radius and contain a number of unbound particles, the latter of which in particular drives up the velocity dispersion.

The velocity anisotropy $\beta$ hovers around zero for most times and radii of most of the decay parameters we consider.  This is unsurprising, since the initial conditions and the decays have a high degree of spherical symmetry in both configuration and velocity space.  The exceptions in $\beta \approx 0$ occur, unsurprisingly, for small $\tau$ and large \vk.  In the case of $\tau \le 1$ Gyr for $\vke = 50$ or $100 \hbox{ km s}^{-1}$, $\beta < 0$ in the outskirts of the halos at late times, indicating that orbits in the outskirts of those halos become tangentially biased.  Radial orbits in the outskirts of galaxies are easiest to strip since their binding energy is lower at fixed radius than more circular orbits.  For large \vk and $\tau$, though, the orbits clearly become radially biased in the outskirts of the halo.  These halos have not settled into a true dynamical equilibrium, and there is a significant net outflow of particles.  This net outflow drives the radial bias in the orbits, and is more significant in the outskirts because any particle which decays onto an unbound orbit in the interior of the halo must pass through the exterior regions.  Moreover, the particles originating in the outskirts are more vulnerable to ejection, so the daughter particles originating in the outskirts will also have a significant radial bias.  The interior regions of the halo are less vulnerable to decay (unless $\vke \gg \vvire$) and the dynamical time is short relative to the decay time, both of which serve to preserve the initial orbital structure deep within the halos.

However, since the dark-matter mass profile may be inferred observationally but the dark-matter velocities are not, we focus our analysis on the mass distribution in halos as a function of decay parameters and time.  Perhaps the most striking feature of Fig.~\ref{fig:rrho} is that
the NFW halo profile, Eq.~\ref{eq:rho}, is clearly not a good
fit to many of the profiles.  In the way in which we have chosen
to display the density profile information, $r\rho$ as a
function of $r$, the profile should be flat for $r/\rse \ll 1$
and be monotonically decreasing for larger $r$.  However, there
is a clear turnover in $r\rho$ in some cases, most notably in
the $\tau \le 1$ Gyr cases, but also at late times for $\tau =
10$ or 50 Gyr and $\vke \ge 100\hbox{ km s}^{-1}$.  

In order to find the goodness of fit of
the NFW profile quantitatively, as well as the best-fit NFW
parameters, we bin the particles radially with respect to the
center of mass of the system, and use two different likelihood
estimation routines: a simple $\chi^2$ model, which should work
in the limit of a large number of particles in each bin; and a
Poisson-based likelihood.  We treat the uncertainty in the
number of particles in each bin as $\sqrt{N}$, where $N$ is the
observed number of particles in a bin.  Instead of using
$\rhose$ and $\rse$ as free parameters in the NFW fit in
Eq.~(\ref{eq:rho}), we use \Mvir~and $c$.  We use 100 bins
between $r_\mathrm{min}$ ($=4$ kpc, the minimum radius at which
numerical relaxation ceases to matter) and $r_\mathrm{max} =
250$ (the approximate virial radius at $t = 0$).  We bin the
particles either linearly or logarithmically in $r$, and obtain
identical fits, as expected.  The minimum $\chi^2$ parameters
are slightly biased relative to those obtained with the Poisson
likelihood function, but not at a significant level.  Thus, we
can use statistics associated with $\chi^2$ to describe the
fits, and define $\chi^2$ per degree of freedom
($\chi^2$/d.o.f.) as our metric for the fit quality.

\begin{figure*}
\begin{centering}
	\includegraphics[angle=270,width=0.47\textwidth]{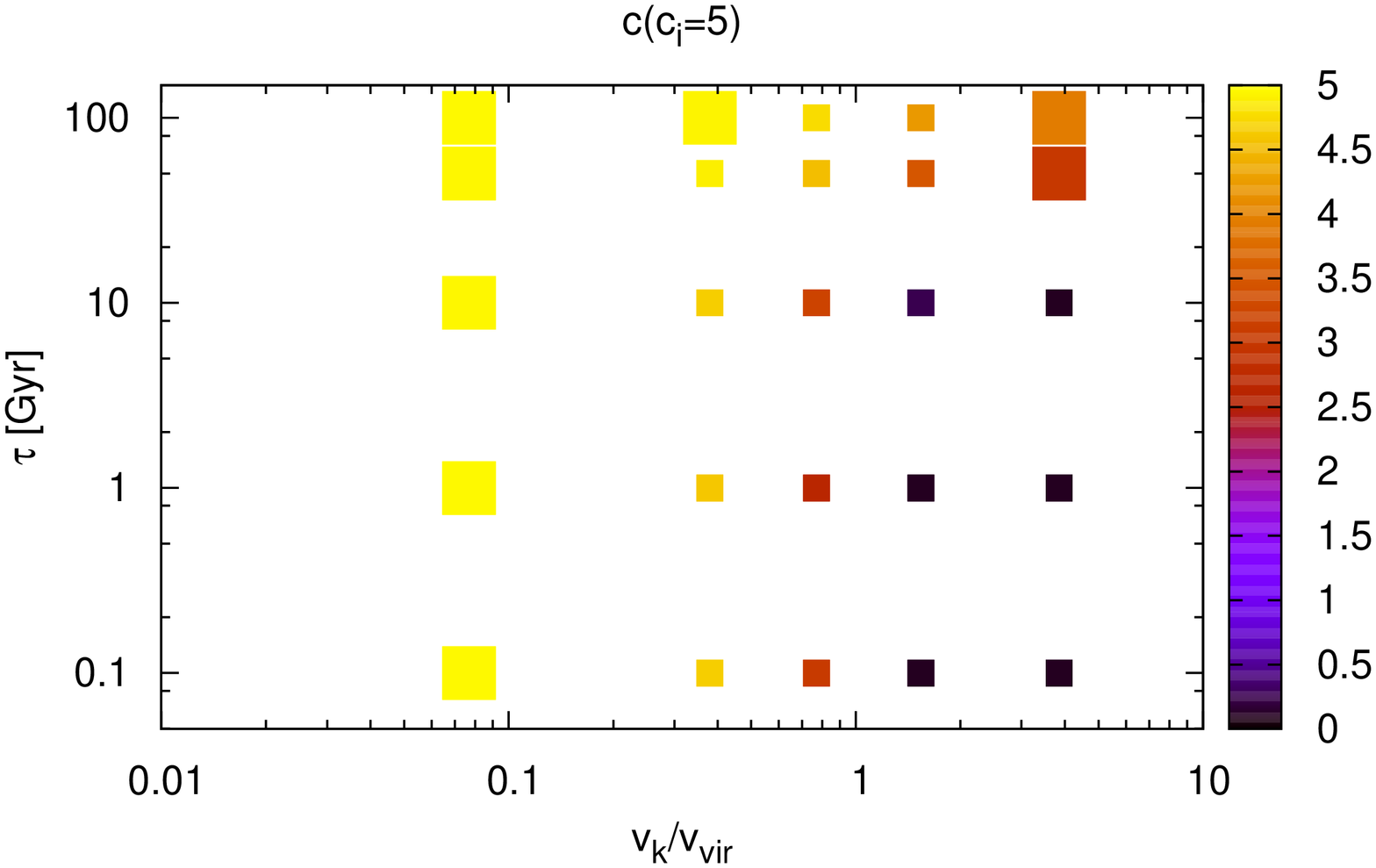} \includegraphics[angle=270,width=0.47\textwidth]{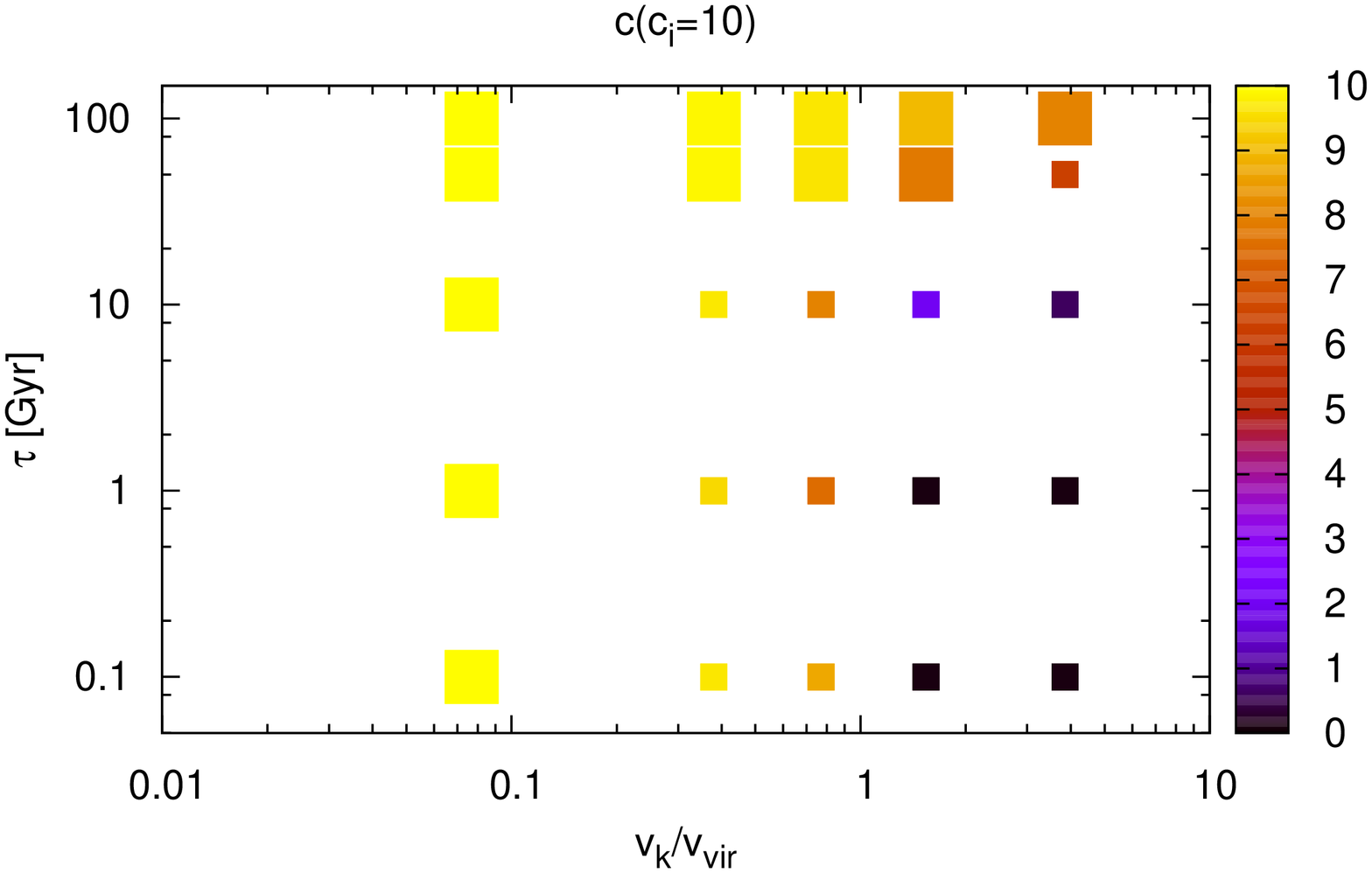}
\end{centering}
\caption{\label{fig:c}Concentration at $t_H$ for an initial $c=5$ (left panel) and $c=10$ (right panel).  The shading indicates the NFW concentration at $t_\mathrm{H}$, and the size of the points indicates the goodness of the NFW fit.  Big points show that the NFW profile is a good fit to the density profile throughout the simulation, and small points indicate significant deviations from NFW fits during the simulations.}
\end{figure*}

We summarize the best-fit NFW concentrations and the
goodness of fit (parametrized by $\chi^2/$d.o.f.) in
Fig.~\ref{fig:c}.  The concentrations are extrapolated to a
Hubble time $t_\mathrm{H} = 13.7$ Gyr from our simulations, and
their values correspond to the shading for each grid point
representing a different ($\vke$,$\tau$) pair used for a
simulation.  The size of the grid points corresponds to the quality of the fit.
If the $\chi^2$/d.o.f. exceeds 1.5 at any time during the
simulation, the grid point is small; if $\chi^2$/d.o.f.$<1.5$
for the whole simulation (sampled every 0.5 Gyr), then the point
is large.  In some cases, the fit is somewhat marginal
$(\chi^2/\hbox{d.o.f.} \sim 1.5)$ much of the time, in the
dividing regions of $\vke-\tau$ parameter space between good and
really bad NFW fits.  The general features between the
simulations with initial $c = 5$ and $c = 10$ are similar,
though; for $\tau \le 10$ Gyr and $\vke/\vvire > 0.4$ ($\vke \ge
50\hbox{ km s}^{-1}$), the NFW profile is generally not a good
fit to the halo profile at some (especially late) times.  For
lower \vk~or high $\tau$, the NFW density profile provides a
reasonable fit throughout the lifetime of a halo, even if the
halo mass and concentration change.  The concentrations drop
rapidly as $\vke$ increases, and the final concentration is less
sensitive, in general, into $\tau$ than to $\vke$.  The
concentration always decreases as a result of decay, which is
expected if kinetic energy is injected to the halo or in the
case of adiabatic expansion.

At large $\tau$ (specifically, $\tau = 50$ and 100 Gyr) and
$\vke = 500 \hbox{ km s}^{-1}$, the changes to the halo
structure with time are well-described by the adiabatic-expansion scheme explored in Ref.~\cite{peter2010}.  If $\tau$
is much greater than the halo dynamical time
($\tdyne \lesssim 1$ Gyr for particles in an $\Mvire =
10^{12}M_\odot$ halo) and the kick speed is high enough that
daughter particles are cleanly ejected from the halo, then
changes to the gravitational potential are adiabatic and easy to model.  If
all particles are on circular orbits, the density
profile remains constant, but with the scale radius
rescaled as $\rse = r_\mathrm{s}(t=0)/(1-f)$ and the NFW scale
density rescaled as $\rhose = (1-f)^4\rhose(t=0)$, where $f$ is
the fraction of the initial dark-matter particles that have
decayed.  We find that NFW profiles are a good fit for $\vke =
500 \hbox{ km s}^{-1}$ ($\vke/\vvire \sim 4$) and $\tau \ge 50$
Gyr with the adiabatic-expansion parameters, but this does
not work so well for $\tau = 10$ Gyr or for $\vke = 200 \hbox{
km s}^{-1}$ ($\vke / \vvire \sim 1.5$).  Thus, it appears that $\tau \gg \tdyne$
and $\vke /\vvire \gtrsim 1$ are required for the analytic adiabatic-expansion description to describe halo evolution.

We reported the goodness of fit and concentration in
terms of $\vke/\vvire$ instead of $\vke$ because
we would like to extrapolate the findings of our
simulations to other halo masses.  It is natural to parametrize
the kick speeds $\vke$ in terms of the virial speed; the virial
speed is a measure of the binding energy of the halo, and $\vke$
is a measure of the kinetic energy injected into the halo.
Following Ref.~\cite{peter2010}, we link the decay time $\tau$
with the crossing time $r/v_\mathrm{circ}$ at the half-mass
radius, where $v_\mathrm{circ} = [GM(r)/r]^{1/2}$, such that the
typical dynamical time of a particle in a halo is
\begin{eqnarray}\label{eq:tdyn}
	\tdyne \approx (G \Deltave \rho_\mathrm{m} )^{-1/2} c^{-3/4}.
\end{eqnarray}
Thus, we can extend our simulation results to halos of other masses.

\begin{figure*}
\begin{centering}
	\includegraphics[angle=270,width=0.47\textwidth]{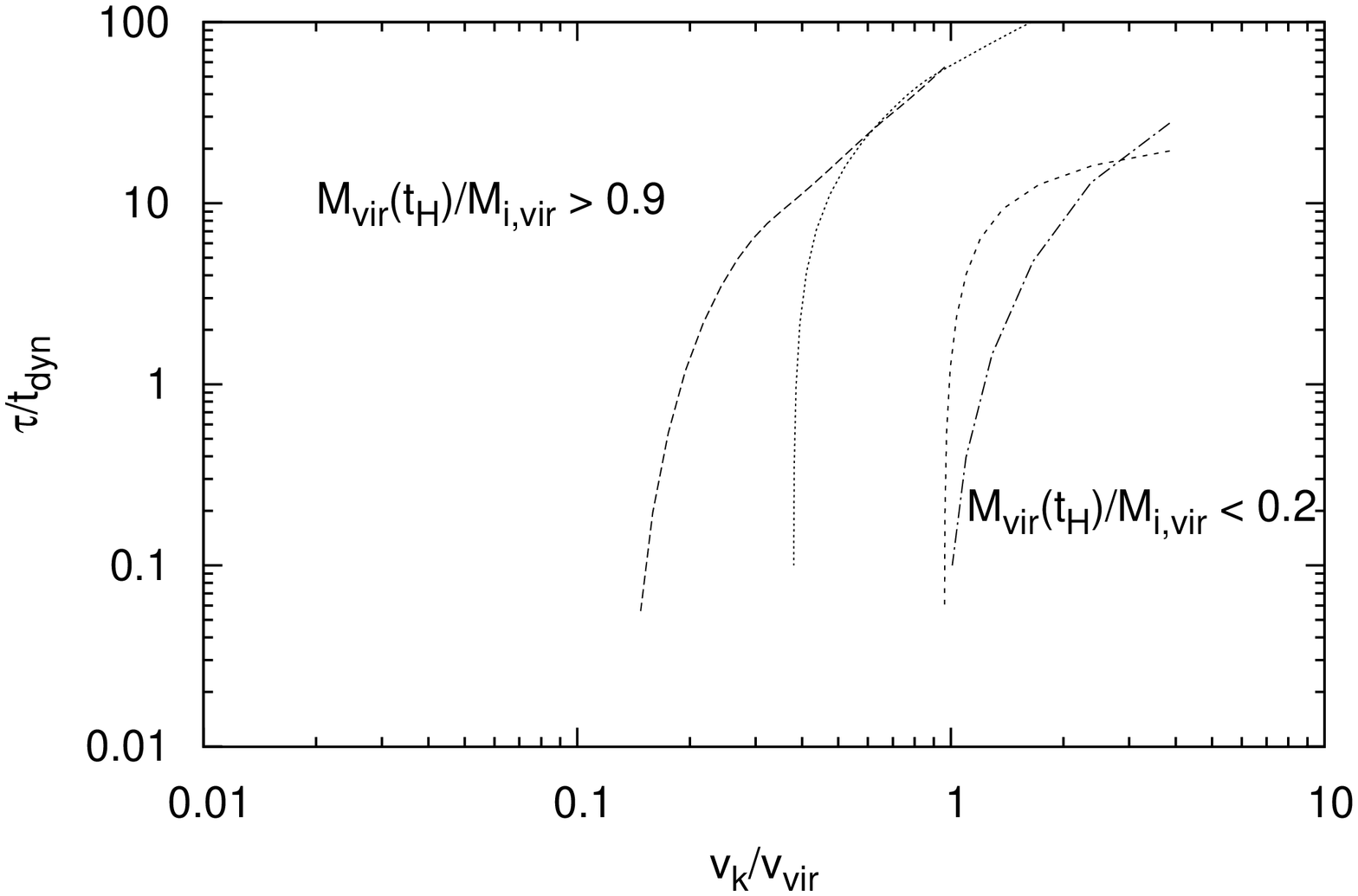} \includegraphics[angle=270,width=0.47\textwidth]{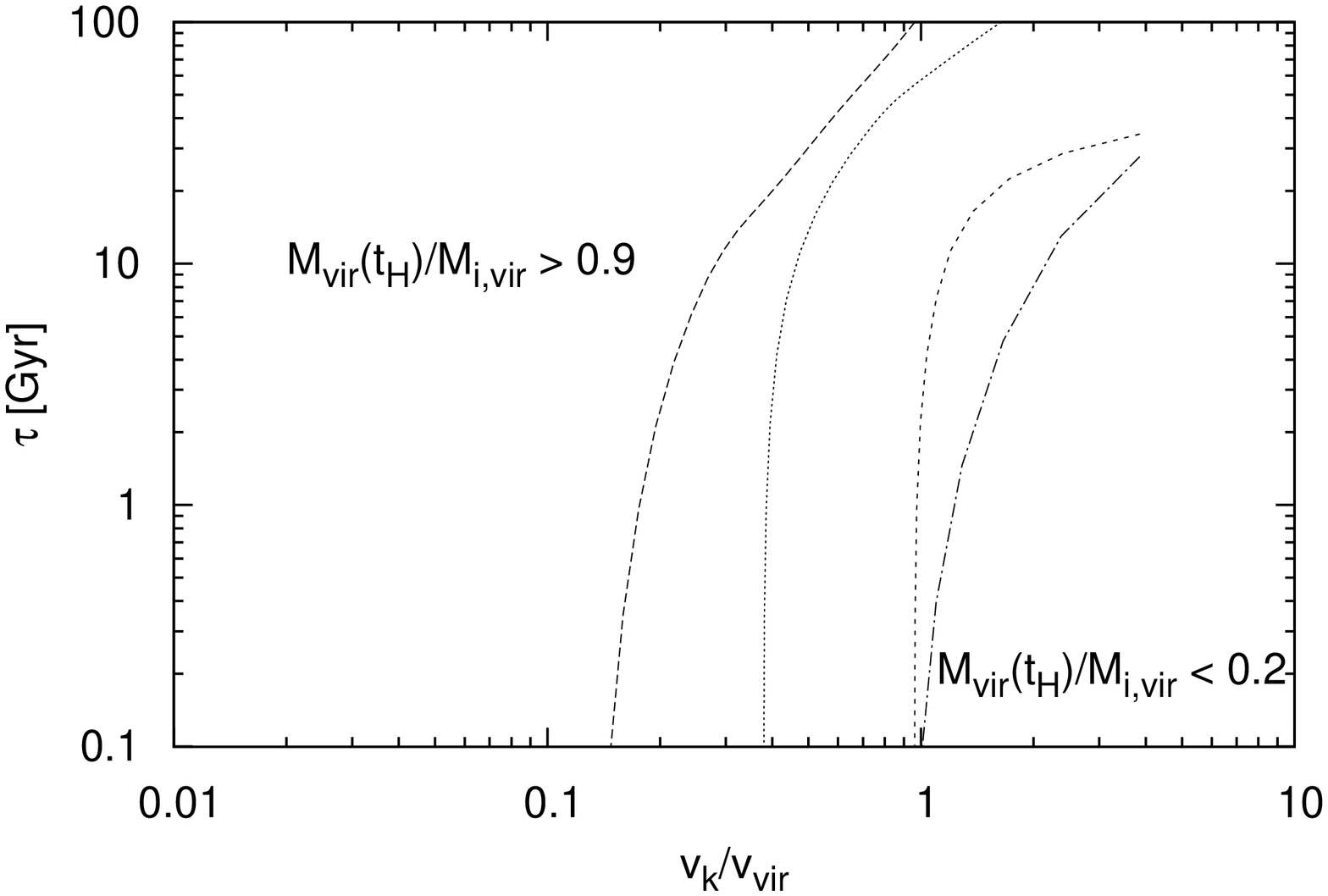}
\end{centering}
\caption{\label{fig:mass}Virial mass remaining at $z=0$ as a function of the initial concentration, $\tau$, and \vk.  The long- and short-dashed lines show where the virial mass after a Hubble time is 0.9 and 0.2, respectively, times the initial virial mass if the initial concentration is $c=5$.  The dotted and dot-dashed lines show the equivalent 0.9 and 0.2 remaining mass fractions for an initial $c=10$. Left panel: Remaining mass as a function of $\tau/\tdyne$ and $\vke/\vvire$.  Right panel: Remaining mass as a function of $\tau$ in Gyr and $\vke/\vvire$.}
\end{figure*}

We summarize the mass remaining after a Hubble time
$t_\mathrm{H}$ in Fig.~\ref{fig:mass}.  In this plot, we show
regions in the decay parameter space in terms of $\vke/\vvire$
and $\tau/\tdyne$ on the left, and $\vke/\vvire$ and $\tau$ on
the right, in order to highlight both the absolute and relative
time scales.  The Hubble time is a fixed time scale, and one can
consider $\tau$ in reference to that timescale, or think of
$t_\mathrm{H}$ in terms of dynamical times, and then consider
the relative time scale $\tau/\tdyne$.  The long-dashed line
shows $\Mvire(t_\mathrm{H})/\Mvire(t=0) = 0.9$ for the $c = 5$
simulations, which has been found via interpolation of our grid
of $\Mvire(t_\mathrm{H})/\Mvire(t=0)$ as a function of the decay
parameters.  This line indicates the transition from decays
having little effect on the virial mass of the halo to having a
nontrivial effect.  The short-dashed line shows
$\Mvire(t_\mathrm{H})/\Mvire(t=0) = 0.2$, which is around the
transition between decays causing total destruction of the halo
and moderate mass loss in the halo.  The dotted and dot-dashed
lines show $\Mvire(t_\mathrm{H})/\Mvire(t=0) = 0.9$ and
$\Mvire(t_\mathrm{H})/\Mvire(t=0) = 0.2$ for the halos which
originally had $c = 10$.

When the decays are parametrized in terms of
$\tau/\tdyne$ and $\vke/\vvire$, the
$\Mvire(t_\mathrm{H})/\Mvire(t=0) = 0.9$ and the
$\Mvire(t_\mathrm{H})/\Mvire(t=0) = 0.2$ lines are largely
similar between the two different initial concentrations.  The
largest discrepancy occurs near $\vke / \vvire \approx 0.2$ and
$\tau < 10$ Gyr.  This effect is real, and is due to the fact
that more concentrated halos have more particles that are more
tightly bound to the halos, which makes the particles harder to eject.
However, for $\tau / \tdyne > 10$ or for the
$\Mvire(t_\mathrm{H})/\Mvire(t = 0) = 0.2$ line, the differences
in mass retention between the two sets of simulations are small.
The differences are greatest at small $\tau/\tdyne$
and $\vke/\vvire$ because the main particles affected and likely
to be kicked out of the halo due to such decays are in the outer
halo.  If $\vke / \vvire \sim 1$,
particles even relatively tightly bound to the halo have a high
chance of being ejected, which ameliorates some differences due
to differences in concentration.  In general, though, even
though the dynamical times of the $c = 5$ and $c = 10$ halos are
different by a factor of 1.7, and even though there are
different numbers of dynamical times in $t_\mathrm{H}$ for the
two sets of halos, the remaining halo mass at $t_\mathrm{H}$ is
not a strong function of concentration in $\tau/\tdyne -
\vke/\vvire$ parameter space.

The differences are more pronounced in the right-hand side plot
in Fig.~\ref{fig:mass}, in which the remaining mass is plotted
as a function of $\tau$ and $\vke/\vvire$.  The $c = 10$ halos
tend to retain more mass at fixed $\vvire$ and $\tau$ because
the typical particle has a higher binding energy than in the
$c=5$ case.  Since the dynamical time depends on $c$, which is
related to the distribution of particle energies within the
halo, some of the dependence of halo mass loss on concentration
is divided out by parameterizing mass loss in terms of $\tau /
\tdyne$ instead of $\tau$.

A more direct comparison would be to examine mass loss and
concentration changes as a function of $\vke/\vvire$ and
$\tau/\tdyne$ for a fixed number of dynamical times.  When we do
this, we find that the $\Mvire(t_\mathrm{H})/\Mvire(t = 0) =
0.2$ lines lie practically on top of each other for $c = 5$ and
$c = 10$, but the $\Mvire(t_\mathrm{H})/\Mvire(t = 0 ) = 0.9$
line is offset by approximately a factor of 2 in $\tau/\tdyne$ between
$c = 5$ and $c = 10$.  However, what matters in comparisons to
observations is the effect of decay at a fixed time.  Instead of
mapping results in terms of $\vke/\vvire$ and $\tau/\tdyne$, we
interpolate in terms of $\vke/\vvire$, $\tau$, and $c$.  Our
simulations span the typical $\Lambda$CDM concentrations
expected for the sizes of halos relevant to the observations,
and the virial speeds span a factor of $\approx 10$, a
relatively small range.

In summary, we find that for halos that initially have $c = 5$
or 10 and have the equilibrium NFW halo profiles expected in
$\Lambda$CDM simulations, the effects of decay in the regime
$\tau/\tdyne \lesssim 10$ and $\vvire/\vke \gtrsim 1$ are to
drastically lower the halo concentration in a Hubble time, drive
substantial mass loss, and drive changes to the halo profile
such that the NFW profile is not a good fit.  For $\tau/\tdyne
\gtrsim 50$ Gyr or $\vke / \vvire \lesssim 0.2$, the NFW profile remains a
good fit to the density profile, mass loss is not too
significant, and the concentration does not drop too much.  Mass
loss and concentration reduction are more severe for fixed
$\tau/\tdyne$ for large $\vke$ until $\vke/\vvire \sim 4$, at
which point the decay can be described analytically in the
adiabatic-expansion approximation, and the effects of decay
no longer depend on $\vke$ as long as $\vke / \vvire \gtrsim 4$.
At such high $\vke$, any daughter particle in a decay is ejected
from the halo, so there is no kinetic energy injection into the
halo as a result of decays.  In the intermediate regime [roughly
corresponding to the space between the
$\Mvire(t_\mathrm{H})/\Mvire(t = 0) = 0.2$ and
$\Mvire(t_\mathrm{H})/\Mvire( t = 0) = 0.9$ lines], mass loss is
moderate although the concentration may drop quite a lot, and the
inner part of the density profile is shallower than for NFW.

There are some differences between the $c = 5$ and $c = 10$
simulations beyond that which can be factored out by
parameterizing decay in terms of the virial time scales
$\vke/\vvire$ and $\tau/\tdyne$.  This is due to the fact that
the gravitational potential, and hence binding energy, does depend
on the concentration in a way that is not completely factored
out by the virial scaling.  Even though we parametrize halos in
terms of ``typical'' dynamical times and typical speeds, there
is a diversity of particle speeds and time scales within a halo.
Nonetheless, since our simulations span the range of \tdyn~and
$c$ of the halo observations below, we are able to interpolate
our simulations and map $\Lambda$CDM dark-matter halos to halos
in a decaying-dark-matter cosmology.

\subsection{Relation to observational constraints}\label{sec:results:obs}
\subsubsection{The mass-concentration relation}

Under the hierarchical picture of structure formation in the
Universe, it is expected that low-mass dark-matter halos form
earlier than high-mass halos.  As a consequence of this early
formation time, the inner parts of low-mass halos are far more
dense relative to the outskirts of the halo relative to
high-mass halos \cite{lu2006}.  In other words, low-mass halos
are more concentrated than high-mass halos.  This trend has been
observed and quantified in simulations of $\Lambda$CDM
cosmologies
\cite{bullock2001,neto2007,maccio2007,duffy2008,maccio2008,dzhao2009}.
It is found that cosmologies with higher matter fractions,
$\Omega_\mathrm{m}$, and a higher amplitude of matter
fluctuations averaged in $8~h^{-1}$ Mpc spheres, $\sigma_8$,
produce higher concentrations for fixed halo mass than
cosmologies in which $\Omega_\mathrm{m}$ and $\sigma_8$ are low.
This is due to the fact that structures collapse earlier when
the Universe is more dense if $\Omega_\mathrm{m}$ and $\sigma_8$
are higher.

\begin{figure*}
	\begin{centering}
	\includegraphics[width=0.95\textwidth]{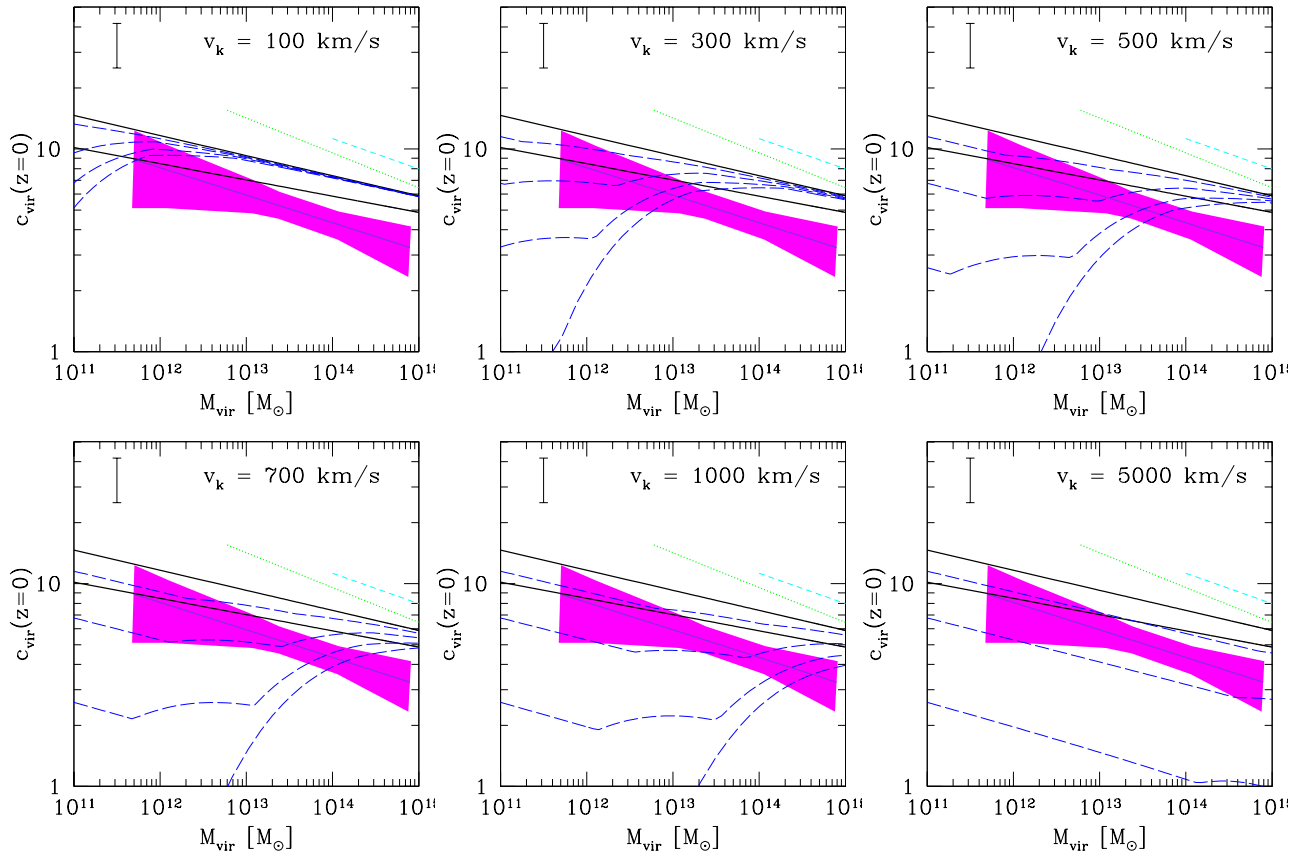}
	\end{centering}
	\caption{\label{fig:mc}The mass-concentration relations predicted for $\Lambda$CDM and decaying-dark-matter cosmologies and inferred from observations of galaxies, galaxy groups, and clusters.  The short-dashed line shows the mean mass-concentration relation found in Ref.~\cite{comerford2006} using cluster observations, and the dotted line shows that from x-ray observations in Ref.~\cite{buote2007}.  The shaded region shows the 1-$\sigma$ fits to the mean mass-concentration relation inferred from weak lensing \cite{mandelbaum2008}.  The upper and lower solid lines show the mean mass-concentration relation found in dissipationless simulations using WMAP-1 and WMAP-3 cosmological parameters, which bracket the WMAP-7 mass-concentration relation \cite{maccio2008}.  The long-dashed lines show the mass-concentration relation for decaying dark matter, with the lines corresponding to (from thickest to thinnest) $\tau = 5, 20, 40,$ and 100 Gyr.  The error bar shows the intrinsic scatter in the theoretical relations, and is also approximately the 1-$\sigma$ error in the mean observationally inferred relations.} 
\end{figure*}

Decays will make halos appear less concentrated; hence,
$\Omega_\mathrm{m}$ and $\sigma_8$ inferred from halos that have
experienced decay \emph{if} one begins with a WIMP dark-matter
ansatz would be lower than that inferred from observations of
earlier epochs.  We can set conservative constraints on the
decay parameter space by comparing observationally inferred
mass-concentration relations with the mass-concentration
relation derived from decays in a high-$\Omega_\mathrm{m}$,
high-$\sigma_8$ cosmology.  The CMB last-scattering surface is
the earliest window into cosmological parameters.  The
six-parameter cosmological fit to the \emph{Wilkinson Microwave
Anisotropy Probe} (WMAP) seven-year data set suggests
$\Omega_\mathrm{m} = 0.267 \pm 0.026$ (1-$\sigma$) and $\sigma_8 =
0.801 \pm 0.030$ (1-$\sigma$) \cite{larson2010}.  The 2-$\sigma$
upper limits of $\Omega_\mathrm{m}$ and $\sigma_8$ correspond
approximately to the mean WMAP first-year mean parameters
\cite{spergel2003}.  The mean mass-concentration relation found
in simulations of a $\Lambda$CDM cosmology using the WMAP-1 parameters is illustrated
in Fig.~\ref{fig:mc}, and is denoted by the upper solid line
\cite{maccio2008}.

We use our grid of mass loss and concentration as a function of
$\tau$ and $\vke / \vvire$, found using the simulations and
described in Sec.~\ref{sec:results:theo}, to map $\Lambda$CDM
halo masses and concentrations in a WMAP-1 cosmology to those of
halos at $t = t_\mathrm{H}$.  We show the mass-concentration
relation as a function of decay parameters in Fig.~\ref{fig:mc},
along with the mean theoretical and observed relations.  It
should be noted that we use the best-fit NFW concentrations
[Eq. (\ref{eq:c})] in Fig.~\ref{fig:mc}, even if the NFW profile
is not a good fit to the density profile.  Nonetheless, the NFW
concentration provides a reasonable measure of the density
contrast within the halo.   

In Fig.~\ref{fig:mc}, the solid lines represent the mean
mass-concentration relations found in dissipationless,
baryon-free cosmological simulations using WMAP one-year (upper
line) \cite{spergel2003} and WMAP three-year (bottom line)
\cite{spergel2007} cosmologies \cite{maccio2008}.  The
long-dashed lines represent decaying-dark-matter models with
(top to bottom) $\tau = 100$, 40, 20, and 5 Gyr under the
assumption that the WMAP one-year cosmology is the underlying
cosmology.  The short-dashed and dotted lines are
observationally inferred mean mass-concentration relations using
a sample of clusters observed in a variety of ways
\cite{comerford2007} and x-ray observations \cite{buote2007},
respectively.  The error bar in the upper left corner of the
plots shows the scatter in the mass-concentration relation for
simulations, and the typical uncertainty in the mean
mass-concentration relation from the observations.  The shaded
region corresponds to the 1-$\sigma$ region of the mean
mass-concentration relation inferred from weak-lensing
observations of Sloan Digital Sky Survey (SDSS) galaxies
\cite{mandelbaum2008}.

We use only the weak-lensing mass-concentration relation to
constrain the decay parameters.  This is because the cluster and
x-ray samples have serious selection effects which have not
been fully quantified.  Reference~\cite{comerford2007} finds that the
concentration of strong-lens clusters is even higher than that
predicted by Ref.~\cite{hennawi2007}, who identified strong-lens
systems in $N$-body simulations.  Dark-matter profiles can only
be inferred from x-ray observations if the halos are relaxed;
relaxed halos have, on average, higher concentration than the
population of halos as a whole \cite{neto2007}.  Thus, until the
selection effects are better quantified for the x-ray and
strong-lens systems, one should not use the
mass concentration inferred from observations of those systems
to constrain dark-matter models.

Even though there are also systematics that have not been fully
quantified in the weak-lensing measurement (see Sec. IIID of
Ref.~\cite{mandelbaum2008} for a sampling of possible effects),
the measurement provides conservative constraints because it
lies slightly below simulation values, on which we base our
decay predictions.  We consider a ($\vke, \tau$) pair of
parameters to be excluded if its mass-concentration relation
lies below the 1-$\sigma$ lower limit of the mean weak-lensing
mass-concentration relation.

Using this criterion, we see from Fig.~\ref{fig:mc} that current
weak-lensing measurements are unable to constrain $\tau$ for
$\vke \lesssim 100\hbox{ km s}^{-1}$.  However, $\tau \lesssim
30$ Gyr is ruled out for $300\hbox{ km s}^{-1} \lesssim \vke
\lesssim 700\hbox{ km s}^{-1}$, and $\tau \lesssim 35-40\hbox{
km s}^{-1}$ for larger values of \vk.  Interestingly, the
constraints from the simulations for large \vk~are nearly
identical to those found analytically using a simple model of
adiabatic expansion (valid only if $\tau/\tdyne \gg 1$ and
$\vke/\vvire\gg 1$) in Ref.~\cite{peter2010}.

We have neglected the effects of baryons on the halo
concentration.  This is largely because it is not clear how
baryons affect the dark-matter halo.  In cosmological
simulations with two different prescriptions for baryon physics,
Rudd et al. (2008) \cite{rudd2008} found either little
difference relative to dark-matter-only simulations, or an
increase in concentration of $\sim 50\%$ above the
dark-matter-only values.  However, even in the latter case, the
minimum allowed $\tau$ for fixed \vk~would only decrease $\sim
15\%-30\%$.

Better constraints may be obtained in the future.  Upcoming deep wide-field surveys will increase the
statistics of weak and strong lensing and x-ray cluster
profiles, and probe down to smaller halo masses
\cite{annis2005,kaiser2002,lsst2009,predehl2006}.  However,
improvements in using the (redshift-dependent)
mass-concentration relation to constrain dark-matter decays (or
cosmological parameters) will require a better understanding of
the systematics and selection effects for the various methods of
measuring the dark-matter halo masses of galaxies and clusters.  This is likely to be a subject of significant future work, since the selection effects are relevant to determining halo mass functions.  The evolution of halo mass functions depends on the growth function.  Since the growth function is sensitive to the equation of state of dark energy, there will be significant efforts to understand any systematic errors that may obscure inferences about dark energy from observations \cite{albrecht2006}.
In addition, there needs to be a better theoretical understanding of
which processes of galaxy evolution drive changes to dark-matter
halo profiles.  This subject has generated recent interest \cite{romano-diaz2008,pedrosa2009,tissera2009,governato2010,duffy2010}, and with continuing improvements in the spatial resolution and star-formation and feedback recipes in simulations, it should be possible to determine which physical processes influence the distribution of dark matter within galaxies.

\subsubsection{Galaxy-cluster mass functions}

The mass function of galaxy clusters $n(>\Mvire)$ is sensitive
to $\Omega_\mathrm{m}$ and $\sigma_8$ for similar reasons as for
the mass-concentration relation; if the amplitude of matter
fluctuations is higher, structure forms earlier and is more
abundant in the local Universe
\cite{press1974,bond1991,white1993,sheth2002,robertson2009}.
Since galaxy clusters are the largest and rarest of virialized
structures, corresponding to small fluctuations in the linear
matter density field smoothed on $\sim 10$ Mpc scales, small
changes in $\sigma_8$ and $\Omega_\mathrm{m}$ will result in
large changes to the cluster mass function.

If dark matter were to decay into relativistic particles, the
cluster mass function would look different than the $\Lambda$CDM
mass function because the continuous pumping of relativistic
particles into the Universe would alter the background evolution
of the Universe and hence the density threshold for halo
collapse.  This effect was investigated by Oguri et al. (2003)
\cite{oguri2003}.  However, the main effect of decays in which
the mass splitting between dark-matter species is small is to
reduce the mass of halos due to ejection resulting from decays.
Since $\sigma_8$ and $\Omega_\mathrm{m}$ should be the same
regardless of the redshift of the matter tracer in a
$\Lambda$CDM universe, the smoking gun for decays would be if
$\Omega_\mathrm{m}$ and $\sigma_8$ inferred from an early epoch
(the cosmic microwave background or Lyman-$\alpha$ forest) were
systematically higher than that inferred from the $z=0$ cluster
mass function.

We use parameter estimates from the WMAP seven-year data set and
low-redshift observations of clusters to constrain decay
parameter space.  As with the analysis using the
mass-concentration relation, we find conservative bounds on the
decay parameter space by considering the parameters for which
the cluster mass function with $\Omega_\mathrm{m}$ and
$\sigma_8$ set to 2-$\sigma$ above the WMAP-7 parameters maps to
a cluster mass function that is 1-$\sigma$ below $\Omega_\mathrm{m}$ and
$\sigma_8$ inferred from observations of low-redshift
clusters.  Both optical and x-ray observations of clusters
indicate that the 1-$\sigma$ lower limits on $\Omega_\mathrm{m}$
and $\sigma_8$ are approximately the same as the 2-$\sigma$
lower limits on those parameters from WMAP-7, $\Omega_\mathrm{m}
\approx 0.2$ and $\sigma_8 \approx 0.7$
\cite{mantz2008,dunkley2009,henry2009,mantz2009a,vikhlinin2009b,rozo2010}.

We use the relation between the $\Lambda$CDM halo masses and the
halo masses after $t=t_\mathrm{H}$ with decay to find the
cluster mass function,
\begin{eqnarray}
	n_\mathrm{f} (> M_\mathrm{f} ) = \int_{M_{\mathrm{f}}}^\infty dM_\mathrm{f}^\prime \frac{dn_\mathrm{i}(M_{\mathrm{i}}(M^\prime_\mathrm{f}))}{dM_\mathrm{i}} \frac{dM_{\mathrm{i}}}{dM_\mathrm{f}^\prime}.
\end{eqnarray}
Here, $n_\mathrm{f}(> M_\mathrm{f})$ is the comoving number
density of halos with masses larger than $M_\mathrm{f}$ after a
Hubble time of decay, $dn_\mathrm{i}/dM_\mathrm{i}$ is the
Tinker et al. (2008) \cite{tinker2008} $\Lambda$CDM differential
mass function, $M_\mathrm{i}$ is the initial $\Lambda$CDM mass
of the halos, and $M_\mathrm{f}$ is the final halo mass after a
Hubble time of decay.  Since the relation between $M_\mathrm{i}$
and $M_\mathrm{f}$ masses depends on concentration
(Sec.~\ref{sec:results:theo}), we use the WMAP-1
mass-concentration relation from Ref.~\cite{maccio2008} to set
the initial concentrations of the initial $\Lambda$CDM halos and
interpolate among our simulations to find $M_\mathrm{f}$ from
$M_\mathrm{i}$.

\begin{figure*}
	\begin{centering}
	\includegraphics[width=0.8\textwidth]{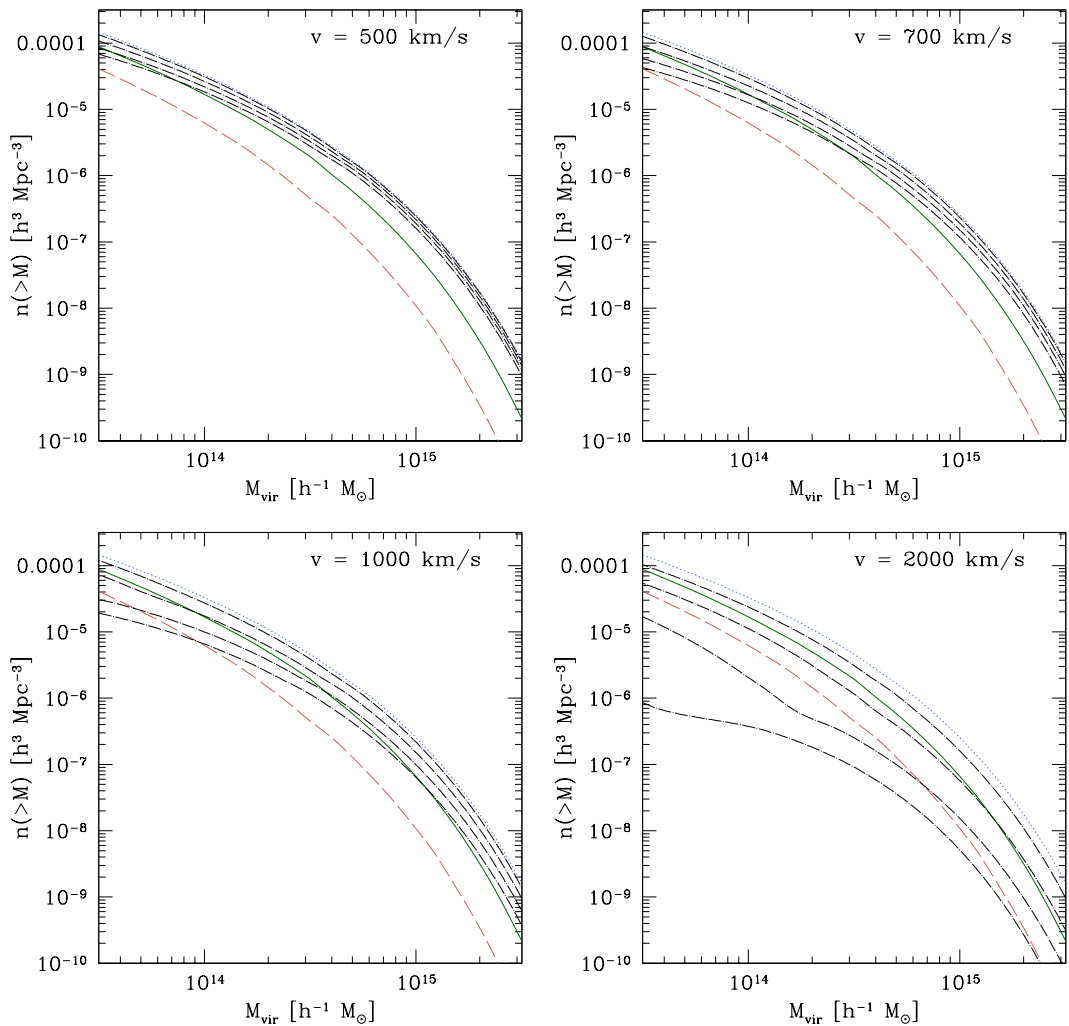}
	\end{centering}
	\caption{\label{fig:dn}z=0 cluster mass functions.  In each panel, the dotted (blue) line represents the Tinker et al. (2008) \cite{tinker2008} mass function for a cosmology with $\Omega_m = 0.318$ and $\sigma_8 = 0.868$, but with all other parameters set to the mean values found in the six-parameter fit of the WMAP-7 data \cite{larson2010}.  The solid (green) line represents the WMAP-7 six-parameter mean cosmology, and the long-dashed (red) line represents a cosmology with $\Omega_m = 0.198$ and $\sigma_8 = 0.724 $.  The short-dashed (black) lines show cluster mass functions for a cosmology with $\Omega_m = 0.318$ and $\sigma_8 = 0.868$, but with decaying dark matter with (bottom to top) $\tau = 5,20,40$, and $100$ Gyr.  For clarity, each panel shows a different \vk.}
\end{figure*}

We show the galaxy-cluster mass function as a function of decay
parameters in Fig.~\ref{fig:dn}.  The dotted line is the mass
function for the high-$\Omega_\mathrm{m}$, high-$\sigma_8$
cosmology, the solid line is for the mean WMAP-7 cosmology, and
the long-dashed line represents the mass function for a
cosmology set to the 2-$\sigma$ lower limits of
$\Omega_\mathrm{m}$ and $\sigma_8$ parameters from WMAP-7 from
the six-parameter cosmological fit \cite{larson2010}.  All other
cosmological parameters have been set to the WMAP-7 mean values,
as those parameters do not affect the normalization or shape of
the mass function nearly as much as $\Omega_\mathrm{m}$ and
$\sigma_8$.  The short-dashed lines show the mass functions
assuming (top to bottom) $\tau = 100$, 40, 20, and 5 Gyr.  Each
panel shows a different \vk.

It is clear that the cluster mass function (for which we
consider $\Mvire \gtrsim 10^{14}M_\odot$ to be classified as a
cluster) is relatively insensitive to small kick speeds, $\vke \lesssim
500\hbox{ km s}^{-1}$.  This is because the typical
virial speed of a cluster is $\vvire \gtrsim 600 \hbox{ km
s}^{-1}$.  From Fig.~\ref{fig:mass}, it is apparent that severe
mass loss in dark-matter halos requires $\vke/\vvire \gtrsim 1$
and $\tau \lesssim 10$ Gyr.  However, the shape of the mass
function is somewhat different than those predicted in
$\Lambda$CDM cosmologies, since higher mass halos are less
affected by lower $\vke$ than low-mass halos.  For $\vke =
1000\hbox{ km s}^{-1}$, the number density of $\Mvire >
10^{14}M_\odot$ is approximately the same for $\tau$ as the
lowest number density allowed by low-redshift observations of
clusters.  However, even for higher $\tau$, the cluster mass
function is far less steep than predicted by $\Lambda$CDM.  At
$\vke = 2000\hbox{ km s}^{-1}$, the cluster number density is
too low for $\tau \lesssim 30$ Gyr.  At even higher $\vke$, the
cluster mass functions converge to the analytic estimates of
Ref.~\cite{peter2010} in the case of $\vke /\vvire \gg 1$, and
constrain $\tau \gtrsim 35-40$ Gyr.  The shape of the cluster
mass function as well as the number density $n(>10^{14}M_\odot)$
can be used to constrain decays, but the statistics are
generally lower for higher cluster masses.

Constraints on dark-matter decays from the cluster mass function
are more robust to baryon physics than the mass-concentration
relation.  Using $N$-body hydrodynamic simulations with two
different models for baryon physics, Rudd et al. (2008)
\cite{rudd2008} found that the normalization of the halo mass
function deviates by $\lesssim 10\%$ with the inclusion of
baryons relative to the halo mass function found in $N$-body
cold-dark-matter-only simulations.

The redshift evolution of the cluster mass function is also a
useful probe of decays~\cite{peter2010}.  The signature of
decays will be that,
for a fixed $z=0$ mass function, the high-redshift mass
functions will have higher normalization and a different shape
than in a $\Lambda$CDM universe.  If one were to estimate
$\Omega_\mathrm{m}$ and $\sigma_8$ for different redshift slices
with the ansatz of a $\Lambda$CDM universe, one would estimate
systematically higher $\sigma_8$ and $\Omega_\mathrm{m}$ at
higher redshifts.  Currently, $\sigma_8$ and $\Omega_\mathrm{m}$
inferred from small samples of x-ray clusters up to $z = 0.7$
are consistent at different redshift slices, although the error
bars are still fairly large
\cite{mantz2008,henry2009,mantz2009a,vikhlinin2009b}.  Upcoming
deep optical, x-ray, and Sunyaev-Zel'dovich surveys
\cite{annis2005,kaiser2002,lsst2009,predehl2006,carlstrom2002,hincks2009,staniszewski2009}
should allow for better constraints on decays, since they will
find large samples of clusters at a variety of redshifts.

\subsection{Summary}

\begin{figure}
	\begin{centering}
		\includegraphics[angle=270,width=0.45\textwidth]{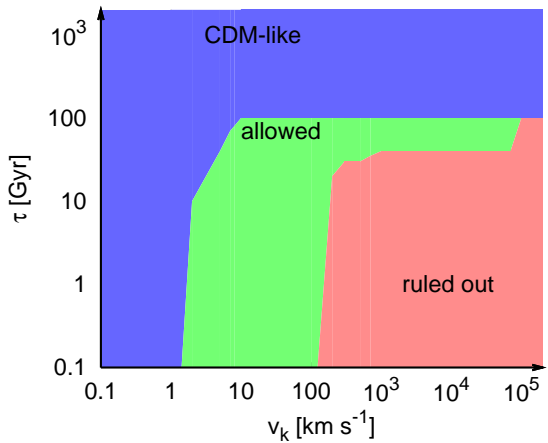}
	\end{centering}
	\caption{\label{fig:summary}Summary of the mass-concentration and $z=0$ cluster mass function limits on the decay parameter space.}
\end{figure}

We summarize the constraints on the decaying-dark-matter
parameter space in Fig.~\ref{fig:summary}.  The exclusion
region, marked ``ruled out'' on the plot, comes from three types
of observations.  For relativistic \vk, the WMAP-1 data set the
strongest constraint on $\tau$, with $\tau \gtrsim 123$ Gyr at
68\% C.L. \cite{ichiki2004}.  This restriction is essentially a
restriction on $\rho_\mathrm{r}(z)$, the radiation density of
the Universe as a function of time.  For $\vke \gtrsim
2000\hbox{ km s}^{-1}$, both the galaxy-cluster mass function
and the mass-concentration relation require $\tau \gtrsim 40$
Gyr.  At lower \vk, the mass-concentration relation sets best
constraints on $\tau$, at approximately the level $\tau \gtrsim
30$ Gyr for $300 \hbox{ km s}^{-1} \lesssim \vke \lesssim
700\hbox{ km s}^{-1}$ and $\tau \gtrsim 20$ Gyr for $100\hbox{
km s}^{-1} \lesssim \vke \lesssim 300\hbox{ km s}^{-1}$.

We note that this exclusion region looks similar to that of
Ref.~\cite{peter2010}.  However, the exclusion region in this
paper is more robust; the exclusion region for $\vke \lesssim
5000\hbox{ km s}^{-1}$ in Ref.~\cite{peter2010} was based on
noting that the clustering of galaxies in the SDSS appeared to
be consistent with the clustering of $\Lambda$CDM halos with $\Mvire \gtrsim
10^{12}M_\odot$ \cite{zehavi2005}.  The constraint was set based
on not letting $\Lambda$CDM halos with $\Mvire \sim
10^{12}M_\odot$ lose more than half their mass due to decays.
This is a fairly hand-waving constraint.  It will likely take cosmological simulations of
decaying dark matter in order to map galaxy clustering to
constraints on dark-matter decays.

The regions labeled ``CDM-like'' (``cold-dark-matter-like", such as WIMPs) and ``allowed'' are still
plausible regions of parameter space.  However, it will be
challenging to probe the region marked ``CDM-like''.  For $\tau
\gtrsim 100$ Gyr, there is hardly any change to halo properties;
mass loss relative to $\Lambda$CDM halos is on the $\sim 10\%$
level, and concentrations go down by at most $\sim 20\%$.  The
NFW profile is still a reasonable fit to the halo density
profiles.  It is not clear if the systematics in even the
largest and deepest all-sky surveys will be small enough to
probe such subtle deviations from $\Lambda$CDM halo properties.
Decay models with $\vke \lesssim 1\hbox{ km s}^{-1}$ will also
be challenging to probe, as such speeds are far less than the
typical virial speeds of galaxies or larger systems.  The
smallest halos that have been found observationally have $\Mvire
\sim 10^9M_\odot$, and are subhalos at that
\cite{strigari_nat2008,vegetti2009c}.  Such halos have $\vvire
\approx 13 \hbox{ km s}^{-1}$.  As we saw in
Sec.~\ref{sec:results:theo}, the structural properties of
dark-matter halos hardly change if $\vke/\vvire \ll 1$.
Moreover, there are poor statistics of such small halos, and it
is not clear how many halos of that size will host a luminous
component.  The statistics of such small objects may be probed
using gravitational milli-lensing, but we are years away from
building any sort of statistical sample of such halos
\cite{koopmans2009,marshall2009,moustakas2009}.

There is some hope of further constraining the region marked
``allowed'' in the near future.  Sunyaev-Zel'dovich, x-ray, and
deep optical surveys should uncover $\sim 10^5$ galaxy clusters
up to quite high redshift
\cite{kaiser2002,annis2005,predehl2006,hincks2009,lsst2009,staniszewski2009};
then, both the $z=0$ cluster mass function and the evolution of
the cluster mass function could be used to more accurately
constrain the decay parameter space.  The deep optical surveys
should also allow for a more precise measurement of the mean
mass-concentration relation, and down to lower halo masses.  On
the theoretical side, a better understanding of selection
effects will allow for
a better comparison of the decaying-dark-matter predictions with
observations.  In addition, if cosmological simulations of
decaying dark matter are performed, one can hope to characterize
the halo occupation distribution (HOD)
\cite{seljak2000,zheng2002,abazajian2005a,zehavi2005,zheng2005,zheng2007},
which will open up galaxy clustering as an avenue to better
explore the decay parameter space.

So far, we have not explored using dwarf galaxies to constrain
the allowed region of parameter space, as suggested by
Ref.~\cite{abdelqader2008}.
This is because there are no robust measurements of the
abundance of $\Mvire \lesssim 10^{11}M_\odot$ halos, nor are
there strong constraints on the inner profiles of dwarf galaxies
\cite{gilmore2007,strigari2007d,strigari2008,walker2009}.\footnote{We
also disagree with several of the technical aspects of
Ref.~\cite{abdelqader2008}: (1) The mechanical kinetic energy
injected by the kick velocity is far more important than the
change in the rest-mass energy they consider. (2) Our analysis
shows that the preservation of a NFW profile, assumed therein,
is not valid.  (3) We find that halos may undergo disruption
even for $\vke\lesssim \vvire$.}  One
could imagine that the $\Mvire \sim 10^9M_\odot$ dwarf galaxies
could be the result of decays of halos born with much higher
virial masses, although the density profile would be
significantly different than expected in $\Lambda$CDM.  In the
future, large optical surveys such as LSST will look farther
down the luminosity function, both in the Local Group and in the
local universe \cite{tollerud2008,lsst2009}, which will allow
for better constraints using the abundance of low-mass halos.
Astrometric observations of stars in Local Group dwarf galaxies
will be critical getting the three-dimensional stellar kinematic
velocity data for better constraints on the dark-matter density
profiles in those galaxies \cite{strigari2007d}.  This will be
possible with next-generation astrometric satellites, such as
\emph{Gaia} and SIM \cite{wilkinson1999,strigari2007d}.

\section{Discussion}\label{sec:discussion}

In the previous section, we presented the results of our
simulations and showed how those results, along with the
mass-concentration relation and cluster mass functions inferred
from existing observations, could be used to constrain the
decaying-dark-matter parameter space.  In this section, we
address two issues that will affect future attempts to further
constrain the decay parameter space: numerical convergence in
simulations, and more accurate mapping of simulations to
observations.

\subsection{Resolution effects}\label{sec:discussion:res}
We based all our analysis in Sec.~\ref{sec:results} on
simulations that initially had $\sim 10^6$ particles within the
virial radius.  We would like to know which of the results we
described above are fully converged, and which results may be
spurious due to resolution effects.  Moreover, we would like to
understand resolution effects before embarking on cosmological
simulations, which are far more costly than the simulations
described in Sec.~\ref{sec:sims}.  If $\gtrsim 10^6$ particles
are required within a virial radius in order to determine halo
properties, it becomes prohibitive to get good statistics on
dark-matter halos (and subhalos) on a variety of mass scales in
cosmological simulations.

In order to understand the robustness of our results, and to
determine if one can get away with fewer particles in the virial
radius for cosmological simulations, we perform simulations with
the same $c$, \vk, and $\tau$ values as those used in
Sec.~\ref{sec:results}, but with $10^5$ particles within the
virial radius.  The low-resolution simulations are briefly described
in Sec.~\ref{sec:sims}.  Our goal is to determine how the virial
mass, concentration, and goodness of fit to the NFW density
profile compare between the two sets of simulations.

\begin{figure}
	\begin{centering}
		\includegraphics[width=0.45\textwidth]{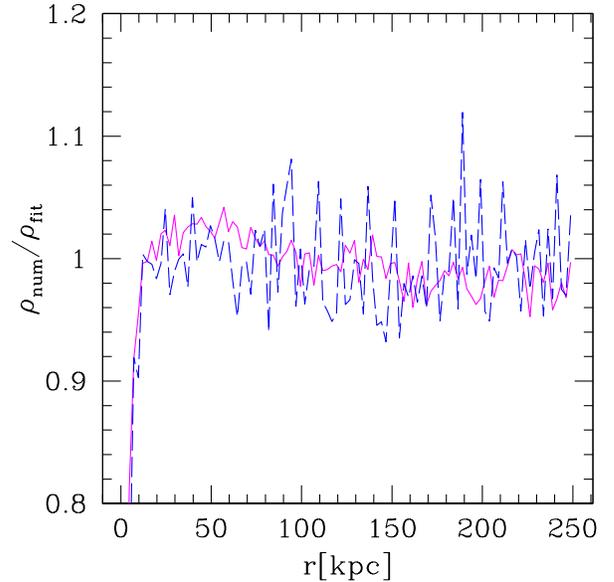}
	\end{centering}
	\caption{\label{fig:res}Density profile for the $c=10$, $\vke = 100\hbox{ km s}^{-1}$, $\tau = 10$ Gyr  simulations at 9 Gyr divided by the best-fit NFW density profile. Solid line: $10^6$ particles in the virial radius.  Dashed line: $10^5$ particles within the virial radius.}
\end{figure}

We find that the virial mass derived from fits to the NFW
profile are consistent between the two sets of simulations, and
are consistent with the virial mass as determined by finding the
average density of dark-matter particles within a sphere
centered on the halo center of mass.  Similarly, the NFW
concentration parameter from the fits is also consistent between
the two sets of simulations.  Therefore, if one is concerned
with accurately determining halo masses or NFW concentration
parameters in a simulation with a decaying-dark-matter
cosmology, one requires no more than $10^5$ particles within the
virial radius.

By contrast, the quality of the NFW fit can be sensitive to the
number of particles within the virial radius.  For example,
$\chi^2 / \hbox{d.o.f.} < 1.4$ for fits to all particles within
the virial radius for $\vke = 100\hbox{ km s}^{-1}$ and $\tau =
10$ Gyr for $10^5$ particles within the virial radius,
regardless of the initial concentration.  However, the $\chi^2$
is bad for $t > 0.5\hbox{ Gyr}$ for both $c=5$ and $c=10$ if
there are initially $10^6$ particles in the virial radius, even
though the NFW fit parameters of the two different-resolution
simulations are identical (see Fig.~\ref{fig:fit} for $\chi^2$
in the $c=10$ case).  This is illustrated in Fig.~\ref{fig:res},
in which we show the density profile from the simulation
($\rho_{\mathrm{num}}$) divided by the best-fit NFW profile for
the two simulations at $t=9$ Gyr with $c=10$, $\vke = 100\hbox{
km s}^{-1}$, and $\tau = 10$ Gyr ($\rho_\mathrm{fit}$).  The
innermost points ($r\lesssim 10$ kpc) show evidence of numerical
relaxation.  For the rest of the halo, one can tell that there
is significant curvature in
$\rho_{\mathrm{num}}/\rho_{\mathrm{fit}}$ for the
$10^6$-particle simulation, but not for the $10^5$-particle
simulations.  This difference is likely in large part due to the
increased Poisson noise in the $10^5$-particle simulation.  In
general, the discrepancy in the quality of the fit appears to be
exacerbated in cases in which $\tau \le 10$ Gyr and $\vke/\vvire
\lesssim 1$.

The consequences for our results are as follows.  Our
constraints on the decay parameter space from cluster mass
functions depended on the mapping between the $\Lambda$CDM halo
mass and the halo mass after a time $t_\mathrm{H}$.  The mass
estimated in the high-resolution simulations used in
Sec.~\ref{sec:results} is virtually identical to that found in
the lower-resolution simulations, so we believe the halo mass
(as a function of decay parameters and time) to be converged and
reliable.  Similarly, the use of the mass-concentration relation
depends on the mapping of $\Lambda$CDM halo masses and NFW
concentrations to those of halos that have experienced decays.
Again, we believe the virial masses and NFW concentrations to be
converged in our simulations.  The shape of the dark-matter
density profile is not used for mapping either of the
observations to constraints on decay parameter space.  Thus, we
believe the mapping of our simulations to observational
constraints to be robust.

The implications for future cosmological $N$-body simulations
are as follows.  We believe it is possible to get accurate
estimates of halo masses and NFW concentrations using no more
than $\sim 10^5$ particles in the virial radius.  It may be
possible to get good statistics using fewer particles, but it
will likely take convergence tests in fully cosmological
$N$-body simulations to determine the minimum particle number
required for convergence, especially for subhalos.  This means
that it may be possible to explore decaying-dark-matter
cosmologies with reasonable statistics using modest
computational resources.  The only situation in which higher
particle numbers may be required is a detailed comparison of the
shape of decaying-dark-matter halo profiles with respect to
$\Lambda$CDM halos.  In this case, we showed that $\sim 10^6$
particles are required within the virial radius to determine the
goodness of the NFW fit, at least down to $\sim 0.02\Rvire$ for
$\vke/\vvire$ and $\tau \lesssim 10$ Gyr.  In other regions of
the decay parameter space, the $10^5$- and $10^6$-particle
simulations converge on the goodness of the NFW fit.  If one is
mainly concerned with radical differences between $\Lambda$CDM
and decaying-dark-matter halo density profiles, then $\sim 10^5$
particles within the virial (or tidal truncation) radius is
likely sufficient; more subtle differences require higher
particle numbers.

\subsection{Mapping simulations to observations}\label{sec:discussion:obs}
In Sec.~\ref{sec:results}, we used NFW fits to the entire halos
to relate halo properties to observables.  However, astronomical
probes of the gravitational potential of galaxies and their
halos are rarely sensitive to the entire halo.  Rotation curves
rarely extend beyond the inner $10\%-20\%$ of the virial radius of
the halo \cite{simon2005,deblok2008,kuzio2008}, strong lensing
is generally also most sensitive to the mass within $\sim
10\%-20\%$ of the virial radius
\cite{koopmans2006,newman2009,richard2009}, and weak lensing is
generally most sensitive to the outer parts of halos (excluding
the inner $\sim 20\%$ of the virial radius)
\cite{mandelbaum2006a,newman2009}. X-ray gas profiles in
galaxies, groups, and clusters can extend up to several tens of
percents from the halo centers
\cite{humphrey2006,gastaldello2007}.  If dark-matter halos were
well described by the NFW profile and there were no systematics
in the observations that would bias the determination of the
dark-matter profile, then all probes should yield the same
profile fits.

Different probes sometimes yield different profile fits, though.
For example, it has been claimed that NFW fits to dwarf-galaxy
rotation curves trend lower in concentrations than predicted in
dissipationless cold-dark-matter simulations, and that NFW
profiles are not always good fits to the data
\cite{kuzio2008,maccio2008}.  Strong-lens systems and x-ray
clusters appear to have higher concentrations than the average concentrations predicted by simulations, although a
large part of that may be due to selection effects
\cite{hennawi2007,mandelbaum2009}.  In some cases, multiple
probes are used on the same system and show different fits.
For example, Ref.~\cite{newman2009} found moderate tension between
the weak-lensing and strong-lensing NFW fits to Abell 611
(although the mass contained within 100 kpc, which is
approximately the transition point between the domains of the
probes, is consistent), and strong tension with NFW fits derived
from stellar kinematic data, so much so that density profiles
significantly shallower than NFW are preferred if the three data
sets are combined \cite{newman2009}.

We would like to explore the possibility that these
discrepancies could be explained in the context of decaying dark
matter.  In Sec.~\ref{sec:results:obs}, we showed that the NFW
profile is not always a good fit to dark-matter halos after some
of the $X$ particles have decayed, especially if $\vke/\vvire
\sim 1$.  Thus, profile fits to subsections of the halo may
deviate from fits to the whole halo.

\begin{figure}
	\begin{centering}
	\includegraphics[width=0.45\textwidth]{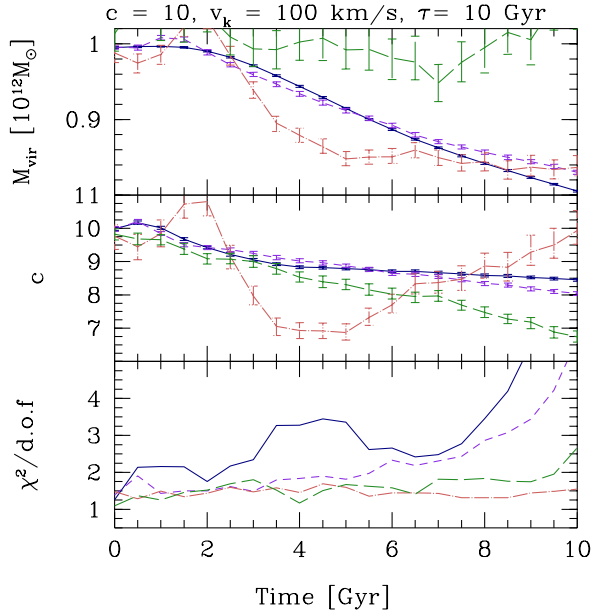}
	\end{centering}
	\caption{\label{fig:fit}NFW fits to a simulation of a $10^{12}M_\odot$, $c=10$ halo with $\vke=100\hbox{ km s}^{-1}$ and $\tau =10$ Gyr.  The solid line represents a fit using all data down to the inner cut-off (4 kpc), the short-dashed line represents fits using particles within half of the initial virial radius, the long-dashed line shows fits within the inner tenth of the initial virial radius, and the dot-dashed line shows fits to the outer half of the halo.  The error bars are the 90\% limits from bootstrap resampling particles at each timestep, although the error bars lose meaning as the $\chi^2$ becomes large. Top panel: NFW fits to the virial mass.  The true virial mass follows the solid line.  Middle panel: The NFW concentration.  Bottom panel: $\chi^2$/d.o.f. for the NFW fit.}
\end{figure}

To demonstrate trends in the fit parameters as a function of
where in the halo one performs the fit, we show in
Fig.~\ref{fig:fit} NFW fit parameters and the goodness of fit
($\chi^2/$d.o.f.) as a function of time for a
halo initially in equilibrium with $\Mvire = 10^{12}\,M_\odot$ and
$c=10$, with $\vke = 100\hbox{ km s}^{-1}$ and $\tau = 10$ Gyr.
The solid (blue) line shows the NFW fit parameters \Mvir~ and
$c$ using all particles within the virial radius.  The error
bars indicate the 90\% range of parameter fits based on
bootstrap resampling particles at each time slice.  The
short-dashed line shows the best-fit parameters using only those
particles in the inner half of the virial radius, and the
long-dashed line shows the fits to particles within the inner
10\% of the virial radius.  In these three cases, the inner
cut-off to the fits is $4$ kpc, corresponding to the minimum
radius at which numerical effects on the density profile can be
ignored.  The dot-dashed line shows the fit resulting from
considering only those particles with radii $>0.5\Rvire$.  In
many cases, the NFW profile does not provide a good fit to the
density profiles, but it is also not always a good fit to
observational data.  The trends we discuss below are there regardless of
the goodness of the NFW fit. 

We find that the concentration derived from the inner parts of
the halo is systematically lower than the concentration derived
from the halo as a whole at late times, although at early times,
the inner concentration is sometimes higher for $\tau \ll
\tdyne$ or $\tau \gg t_\mathrm{H}$.  However, if the maximum
radius used in the fit is $\sim 0.1\Rvire$, the concentration
from the NFW fit is always lower than the concentration from the
NFW fit to the halo as a whole.  In general, the discrepancy
among concentrations becomes less severe as a larger fraction of
the halo particles are used for the fit.  The NFW fit becomes
better and concentrations tend to be slightly higher if the
minimum radius is raised to 10 kpc from 4 kpc, and in general,
NFW fits become better as the maximum radius considered in the
fit decreases.  Fits to the outer part of the halo show a trend
that the halo concentration is initially higher than that
estimated from the halo as a whole, then dips low, and then
becomes relatively high again.  This phenomenon is due to the
finite time it takes for particles ejected from the inner halo
to reach the outer halo, and the longer dynamical (and hence,
equilibration) time in the outer halo.  The NFW profile is
almost always a reasonable fit to the data with $r > 0.5
\Rvire$.

Are these trends consistent with observations?  The relatively
low concentration measured in the innermost part of the halo
relative to the halo as a whole is qualitatively consistent with
the low concentrations inferred from low-surface-brightness and
dwarf-galaxy rotation curves.  However, in the absence of
selection biases, one would expect the weak-lensing-derived
concentrations to be slightly higher than strong-lensing-derived
concentrations for the same halos, and also slightly higher than
concentrations inferred from x-ray gas profiles, which is the
opposite of what is observed.  X-ray and strong-lensing halo
samples are likely highly biased towards higher concentrations,
and so a better understanding of selection effects is necessary
in order to understand the differences in inferred concentration
among the various lensing, x-ray, and dynamical probes of
dark-matter density profiles.

Nonetheless, the prediction from decaying-dark-matter models is
that it is possible that different probes of dark-matter halo
profiles will indicate different concentrations for the same
halo, even if the NFW profile is shown to be a good fit to data
from probes sensitive to the innermost or outermost parts of the
halo, such as weak lensing or rotation curves.

\section{Conclusion}\label{sec:conclusion}
In this work, we have used simulations of two-body dark-matter
decay in isolated halos initially in equilibrium to characterize
the response of halos as a function of the center-of-mass kick
speed \vk, the decay time $\tau$, and time.  We find that for
the union of the parameter space $\vke/\vvire \lesssim 0.2$ and
$\tau/\tdyne \gtrsim 100$, dark-matter halos are essentially
unchanged; the mass loss is $\lesssim 10\%$ over a Hubble time,
and the shape of the density profile does not change, although
the concentration declines somewhat for $\vke/\vvire > 1$.  For
the union of $\vke/\vvire \gtrsim 1$ and $\tau/\tdyne \lesssim
10$, there is severe mass loss and a radical change to the shape
of the dark-matter density profile.  For the rest of the
decaying-dark-matter parameter space, there is moderate mass
loss and significant deviations of the shape of the dark-matter
density profile relative to the initial halo properties.

We used our simulations in conjunction with the observed
mass-concentration relation and the galaxy-cluster mass function
to constrain the two-parameter decay parameter space.  We find
that the mass-concentration relation yields the stronger
constraint on $\tau$ for a broad range of \vk~, constraining
$\tau \gtrsim 30-40$ Gyr for $\vke \gtrsim 200 \hbox{ km
s}^{-1}$.  Constraints on the decay parameters should improve
greatly in the next 5-10 years with blossoming wide-field
Sunyaev-Zel'dovich and optical surveys, and with cosmological
simulations of decaying dark matter. \newline

\vskip 1.5cm

\begin{acknowledgments}We thank Andrew Benson for helpful discussions.  This work was
supported at Caltech by DOE Grant No. DE-FG03-92-ER40701 and the Gordon
and Betty Moore Foundation.
\end{acknowledgments}


\end{document}